\documentclass[final,5p,twocolumn,times,11pt]{elsarticle}
\usepackage{graphicx,color,amsmath,amssymb}
\usepackage{bm}
\usepackage[colorlinks=true, pdfstartview=FitV, linkcolor=red, citecolor=blue, urlcolor=blue]{hyperref}

\advance\voffset -0.2in
\biboptions{sort&compress}

\begin{document}

\begin{frontmatter}
\title{Long-range azimuthal correlations in proton-proton and proton-nucleus collisions from 
the incoherent scattering of partons
}

\author[gm]{Guo-Liang Ma}
\ead{glma@sinap.ac.cn}

\author[ab]{Adam Bzdak}
\ead{abzdak@quark.phy.bnl.gov}

\address[gm]{Shanghai Institute of Applied Physics, Chinese
Academy of Sciences, Shanghai 201800, China}

\address[ab]{RIKEN BNL Research Center, Brookhaven National Laboratory, 
Upton, NY 11973, USA}

\begin{abstract}
We show that the incoherent elastic scattering of partons, as present in a multi-phase transport model (AMPT), with 
a modest parton-parton cross-section of $\sigma=1.5 - 3$ mb, naturally explains the long-range two-particle azimuthal correlation as observed in proton-proton and proton-nucleus collisions at the Large Hadron Collider.
\end{abstract}
\end{frontmatter}

\section{Introduction}
\label{sec: introduction}

Recent experimental observations of the long-range azimuthal correlations in
high-multiplicity proton-proton (p+p) \cite{Khachatryan:2010gv} and
proton-nucleus (p+A) collisions \cite%
{CMS:2012qk,Abelev:2012ola,Aad:2012gla,Adare:2013piz} shed some new light on
our understanding of {\it{fireballs}} created in such interactions.

The measured two-particle correlation function as a function of the
pseudorapidity separation, $\Delta \eta =\eta _{1}-\eta _{2}$, and the
relative azimuthal angle, $\Delta \phi =\phi _{1}-\phi _{2}$, of two
particles demonstrates a great deal of similarity to that measured in peripheral heavy-ion collisions \cite%
{Chatrchyan:2013nka}. In particular, two particles separated by many units
of pseudorapidity prefer to have similar azimuthal angles thus the
two-particle correlation function is peaked at $\Delta \phi =0$. Exactly the
same phenomenon was observed in heavy-ion collisions where it is believed to
originate from hydrodynamical evolution present in such interactions \cite%
{Florkowski:book}. In this picture the initial anisotropic distribution of
matter, characterized e.g. by ellipticity, is translated to the final momentum
anisotropy with $\cos (2\Delta \phi )$ term (and higher harmonics) in the
correlation function. However, the applicability of hydrodynamics to small
systems, as the ones created in p+p and p+A interactions, is questionable
and so far there is no consensus in this matter. Nevertheless, hydrodynamics%
\footnote{%
It should be noted that the long-range rapidity structure is put by hand into
hydrodynamic calculations.} applied to p+p and p+A collisions results in
qualitative and partly quantitative understanding of various sets of data %
\cite%
{Bozek:2011if,Shuryak:2013ke,Bozek:2013uha,Bzdak:2013zma,Qin:2013bha,Werner:2013tya,Bozek:2013ska}%
. On the other hand, the Color Glass Condensate \cite{Gelis:2010nm}, the
effective description of low-x gluons in the hadronic/nuclear wave function,
results in equally good description of the two-particle correlation functions %
\cite{Dusling:2013oia} (see also \cite{Kovchegov:2012nd,Kovner:2012jm} for a more qualitative discussion). 
The advantage of the CGC approach over hydrodynamics is its microscopic character
and internal consistency. On the other hand, hydrodynamics naturally
describes various sets of data for which the CGC predictions are often not 
clear. Moreover, hydrodynamics provides a solid intuitive understanding of the observed signal which is not the case for the CGC. To summarize, at present we have two competing languages\footnote{%
In Ref. \cite{Gelis:2013rba} both physical pictures are argued to be rather connected.} to understand small systems and it is crucial to
establish the true origin of the long-range azimuthal correlation. Several
observables and arguments \cite%
{Bzdak:2013zla,Bozek:2013sda,Coleman-Smith:2013rla,Bjorken:2013boa,McLerran:2013oju,Rezaeian:2013woa,Basar:2013hea, Bzdak:2013rya,Yan:2013laa,Bzdak:2013raa,Konchakovski:2014wqa,Bzdak:2013lva,Sickles:2013yna,Noronha:2014vva}
were recently put forward which hopefully can help to resolve this interesting
issue.

In this paper, we calculate the two-particle density function, $N^{\mathrm{%
pair}}(\Delta \eta ,\Delta \phi )$, in p+p and p+Pb collisions assuming the
incoherent elastic scattering of partons, as present in a multi-phase transport
model (AMPT) \cite{Lin:2004en}. This approach is simple and intuitive, and
more importantly is closely related to quantum chromodynamics (QCD). The
cascade model with the reasonable parton-parton cross-section, $\sigma =1-10$
mb, was proved to be very successful in understanding many features of heavy-ion
collision data, see e.g. \cite{Ma:2010dv,Xu:2011fi,Solanki:2012ne,Ma:2013pha}. This
approach has one crucial advantage over hydrodynamics, namely, there is no
need to assume local thermalization. So far such a calculation was not
published and it is important to establish whether a simple incoherent
scattering of partons with a reasonable partonic cross-section can generate the long-range structure in p+p and p+A two-particle correlation
functions.\footnote{%
We note that the negative result was reported by the CMS Collaboration in
Ref. \cite{CMS:2012qk}. Our results contradict their conclusion.}

Our main result is that the incoherent elastic scattering of partons, with a
partonic cross-section of $\sigma =1.5 - 3$ mb, naturally generates the
long-rage azimuthal correlation of charged particles both in p+p and p+A collisions.
A near side peak at $\Delta \phi =0$ grows with the growing number of
produced particles due to the growing density of partons, and consequently
the larger number of partonic scatterings. The $p_{T}$ dependence of the
near-side peak is also reproduced that is, the signal at $\Delta \phi =0$ is
best visible for $1<p_{T}<2$ GeV/$c$.

In the next section we give a brief introduction to the AMPT model. In Section \ref{sec:results} we present our results for the two-particle correlation functions in p+p and p+A collisions for various multiplicity and $p_{T}$
bins. We finish our paper with comments in Section~\ref{sec:comments} and conclusions in Section~\ref{sec:conclusions}.

\section{Model}
\label{sec:model}

The AMPT model with string melting mechanism is employed in this work (for comparison we also show some results obtained in the default model). It is initialized with a spatial and momentum distribution of minijet partons and soft string excitations from the HIJING model~\cite{Wang:1991hta}. The string melting mechanism converts all excited strings into quarks and antiquarks according to the flavor and spin structures of their valence quarks (in contrast to the default AMPT model, where only partons from minijets are present). The evolution of a quark-antiquark plasma\footnote{In our context we only need partonic scatterings and the composition of the partonic matter is less important.} is modeled by a simple parton cascade. At present, the parton cascade includes only two-body elastic scatterings with a cross-section obtained from the pQCD with a screening mass~\cite{Zhang:1997ej}. Clearly this is a simplified picture however, we believe it captures the main 
features of parton dynamics present at the early stage of a collision. The parton cascade is
followed by the hadronization, where quarks are recombined into hadrons via a simple coalescence model. 
Finally dynamics of the subsequent hadronic matter is described by a relativistic transport model~\cite{Li:1995pra}.
For more details on the AMPT model we refer the reader to Ref. \cite{Lin:2004en}. 
The recent AMPT studies show that the partonic cross-section of $1.5$ mb can describe many experimental observables at the LHC~\cite{Xu:2011fi,Xu:2011fe,Ma:2013bia,Ma:2013pha}. In particular it was found that the long-range azimuthal correlation can be produced by the parton scatterings in Pb+Pb collisions at $\sqrt{s}=2.76$ TeV~\cite{Xu:2011jm}.

\section{Results}
\label{sec:results}

\begin{figure}[t]
\begin{center}
\includegraphics[scale=0.44]{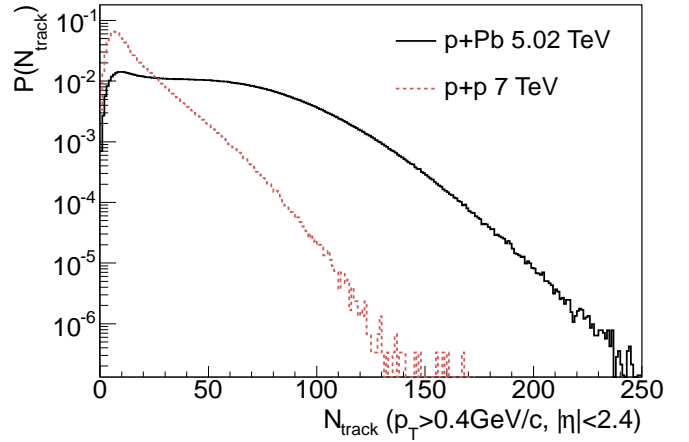}
\end{center}
\par
\vspace{-5mm}
\caption{The multiplicity distribution calculated in AMPT, $P(N_{\mathrm{track}})$, as a function of the number of produced particles, $N_{\mathrm{track}}$, in p+p collisions at $\protect\sqrt{s}=7$ TeV, and p+Pb collisions at $\protect\sqrt{s}=5.02$ TeV for charged particles produced in $|\protect\eta |<2.4$ and $p_{T}>0.4$ GeV/$c$. }
\label{fig:P_Ntrack}
\end{figure}

To directly compare our results with the CMS data we select events with
different values of the number of produced charged particles, $N_{\mathrm{track}}$.
In Figure \ref{fig:P_Ntrack} we present the multiplicity distributions, $P(N_{%
\mathrm{track}})$, in p+p collisions at $\sqrt{s}=7$ TeV and p+Pb interactions at $\sqrt{s}=5.02$
TeV, for charged particles produced in $|\eta |<2.4$ and $p_{T}>0.4$ GeV/$c$. Both
multiplicity distributions are in reasonable agreement with the CMS data%
\footnote{%
We do not compare directly with the CMS data since their $N_{\mathrm{track}%
}^{\mathrm{offline}}$ is not exactly our $N_{\mathrm{track}}$.}, see e.g. %
\cite{talk}.

\begin{figure}
\begin{center}
\includegraphics[scale=0.4]{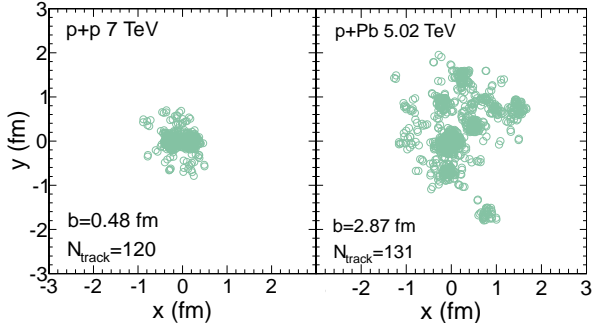}
\end{center}
\par
\vspace{-5mm}
\caption{The initial parton distribution in a p+p collision (left panel) and a p+Pb collision (right panel) for two typical AMPT events (with string melting mechanism) with the number of produced charged particles, $N_{\mathrm{track}}$, larger than $110$ ($|\protect\eta |<2.4$, $p_{T}>0.4$ GeV/$c$). Here $b$ is the impact parameter.}
\label{fig:contour}
\end{figure}

Before we present our main results it could be pedagogical to illustrate the
initial parton distribution in the transverse plane in p+p and p+A
collisions with $N_{\mathrm{track}}>110$. 
As seen in Figure \ref{fig:contour} the initial size of a system in p+p is roughly a factor of $2$ smaller than that in p+A. We checked that in a p+p collision partons are produced mainly in the overlap region of the two colliding protons, leading to a characteristic elliptical shape in a typical p+p event. In a p+A collision, the produced partons are localized in a few spots corresponding to the positions of the wounded nucleons \cite{Bialas:1976ed}.

\begin{figure}[h]
\begin{center}
\includegraphics[scale=0.44]{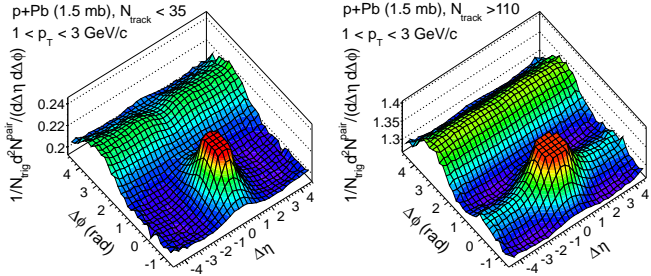}
\end{center}
\par
\vspace{-5mm}
\caption{The AMPT two-particle density function in p+Pb collisions at $\protect\sqrt{s}=5.02$ TeV for low- (left) and high- (right) multiplicity events. The long-range near-side structure in pseudorapidity is clearly visible for high-multiplicity events.}
\label{fig:3D}
\end{figure}

In Fig. \ref{fig:3D} we show the AMPT results for the two-particle density function in p+Pb collisions at $\sqrt{s}=5.02$ TeV as a function of the relative azimuthal angle, $\Delta \phi =\phi _{1}-\phi _{2}$, and the pseudorapidity separation, $\Delta \eta =\eta _{1}-\eta _{2}$, for events with $N_{\mathrm{track}}<35$ (left) and $N_{\mathrm{track}}>110$ (right). In this plot we take the pairs of charged particles with $1<p_{T}<3$ GeV/$c$. In qualitative agreement with the experimental data, the long-range near-side structure is absent for events with $N_{\mathrm{track}}<35$ and is clearly visible in events with $N_{\mathrm{track}}>110$.

\begin{figure*}
\begin{center}
\includegraphics[scale=0.8]{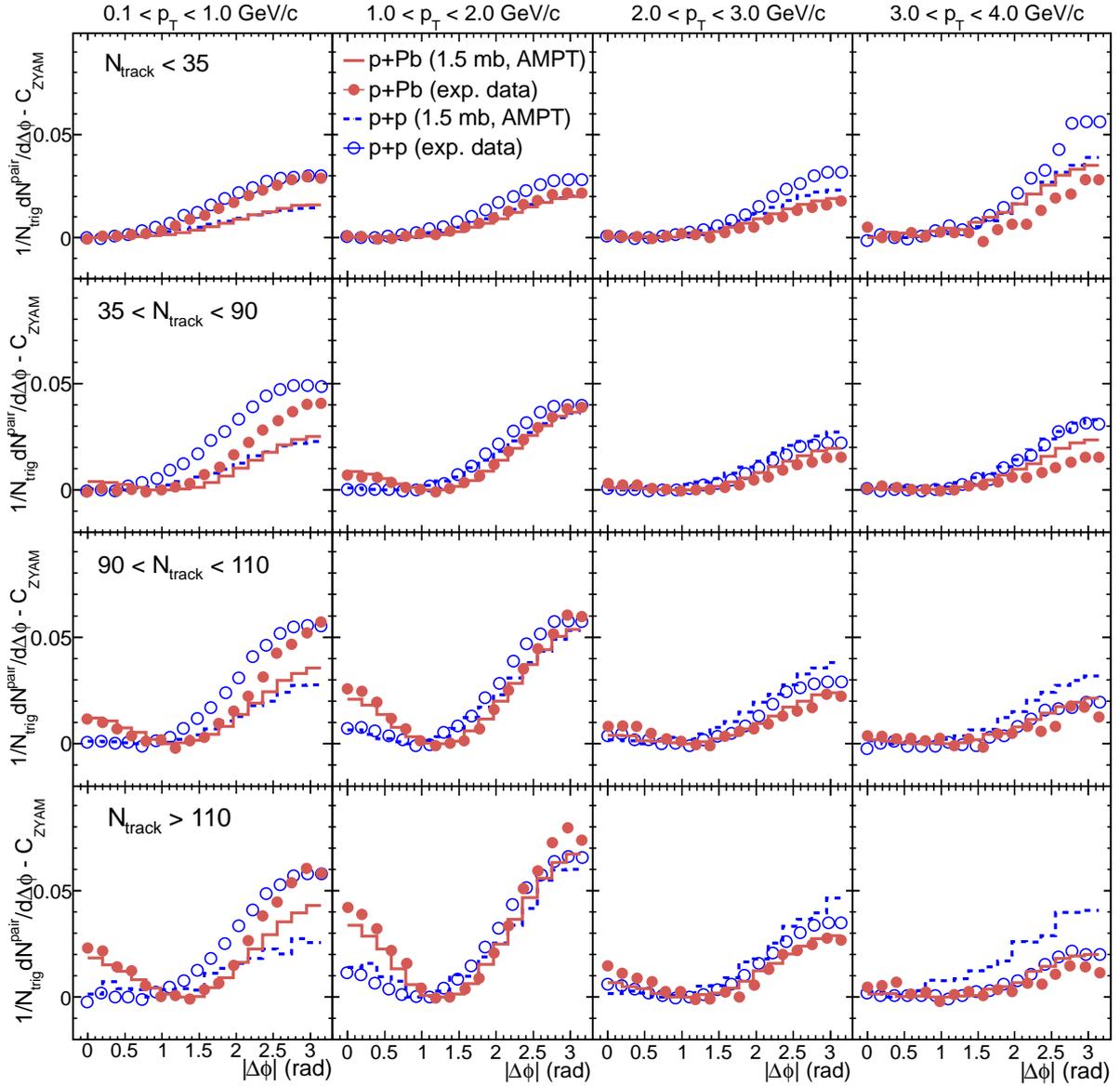}
\end{center}
\par
\vspace{-5mm}
\caption{Distribution of pairs in p+p collisions at $\protect\sqrt{s}=7$ TeV and p+Pb collisions at $\protect\sqrt{s}=5.02$ TeV as a function of the relative azimuthal angle $\Delta \protect\phi $ averaged over $2<|\Delta \protect\eta |<4$ in different $p_{T}$ and $N_{\mathrm{track}}$ bins. Our results (solid and dashed curves) based on the AMPT model (with string melting, $\sigma=1.5$ mb) are compared to the CMS data (full and open circles).}
\label{fig:pPb_main}
\end{figure*}

To compare directly with the data, in Fig. \ref{fig:pPb_main} we present the two-particle distribution functions for
p+Pb collisions at $\sqrt{s}=5.02$ TeV and p+p at $\sqrt{s}=7$ TeV, as a function of the relative
azimuthal angle $\Delta \phi $ and averaged over pseudorapidity region $%
2<|\Delta \eta |<4$%
\begin{equation}
\frac{1}{N_{\mathrm{trig}}}\frac{d^{2}N^{\mathrm{pair}}}{d\Delta \phi }=%
\frac{1}{4}\int_{2<|\Delta \eta |<4}\frac{1}{N_{\mathrm{trig}}}\frac{d^{2}N^{%
\mathrm{pair}}}{d\Delta \phi d\Delta \eta }d\Delta \eta ,
\end{equation}%
for various ranges of $N_{\mathrm{track}}$ and different $p_{T}$ bins. Following the experimental procedure the zero-yield-at-minimum (ZYAM) method is implemented to remove a constant background, $C_{\mathrm{ZYAM}}$. In this calculation we take the partonic cross-section to be $\sigma =1.5$ mb. The AMPT results (solid and dashed curves) are in very good
agreement with the CMS data (full and open circles) for the near-side peak, $\Delta \phi \approx 0$. The agreement with the away-side peak, $\Delta \phi \approx \pi $, is less impressive however, this region is heavily populated
by jets which are of lesser interest in the present investigation. It is worth noticing that at the same $N_{\mathrm{track}}$ bin, the signal at $\Delta \phi =0$ in p+p collisions is noticeably smaller than that in p+A interactions. This feature agrees very well with the CMS data. 

\begin{figure}
\begin{center}
\includegraphics[scale=0.45]{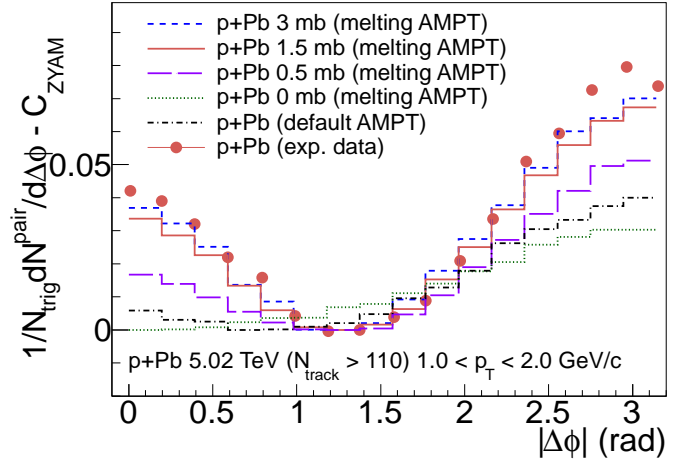}
\end{center}
\par
\vspace{-5mm}
\caption{Distribution of pairs for various values of the partonic cross-section, $\sigma$, in p+Pb collisions at $\protect\sqrt{s}=5.02$ TeV as a function of the relative azimuthal angle $\Delta \protect\phi $ averaged over $2<|\Delta \protect\eta |<4$ for $N_{\mathrm{track}}>110$ and $1<p_{T}<2$ GeV/$c$. Our results (curves) from different AMPT model settings are compared with the CMS data (points). In the default AMPT model only few partons from minijets interact which is not sufficient to produce a sizable signal. In the string melting version all soft strings are converted into partons.}
\label{fig:pPb_main_2}
\end{figure}

In Figure \ref{fig:pPb_main_2} we present the results for p+Pb collisions calculated in the AMPT model 
with various values of $\sigma = 0, 0.5, 1.5$, and $3$ mb. We also show the result of the default AMPT model, where only partons from minijets interact and all soft strings decay independently into particles. In this scenario the number of interacting partons is not sufficiently high to produce a visible effect. On the contrary, in the string melting scenario (in which all initial soft strings melt into partons) the number of interacting partons is significantly larger, roughly a factor of $5$, thus allowing to obtain a sizable signal. As seen in Figure \ref{fig:pPb_main_2} the strength of the signal gradually increases with growing $\sigma$ and, as expected, the signal vanishes completely for $\sigma=0$ mb. It clearly  demonstrates that in the AMPT model partonic scatterings are directly responsible for the signal at $\Delta\phi=0$, as observed in Figures \ref{fig:pPb_main} and \ref{fig:pPb_main_2}.

\begin{figure*}
\begin{center}
\includegraphics[scale=0.85]{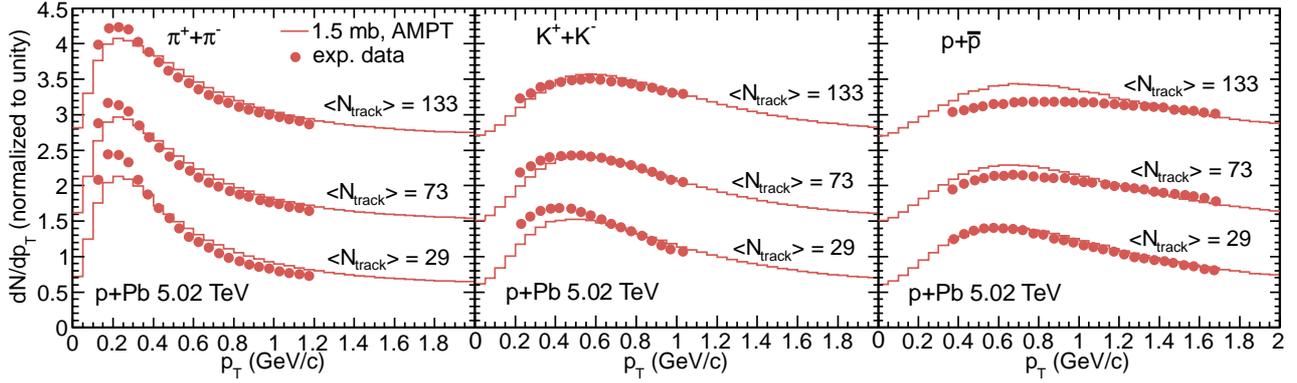}
\end{center}
\par
\vspace{-5mm}
\caption{The transverse momentum spectra (normalized to unity) in $|y|<1$ of produced pions, kaons and protons in p+Pb collisions at $\sqrt{s}=5.02$ TeV for three different centrality classes. The AMPT model (string melting) results are compared to the CMS data (full points). For better visibility, the results for $\langle N_{\mathrm{track}} \rangle_{p_T>0.4 \text{ GeV}/c} = 29$, $73$ and $133$ are shifted vertically by 0.6, 1.5 and 2.7 units, respectively.}
\label{fig:spectra}
\end{figure*}

In the last part of the paper we address the problem of the $p_{T}$ particle spectra. The measured $p_{T}$ distributions evolve towards higher $p_{T}$ with an increasing number of produced particles \cite{Chatrchyan:2013eya}. In principle this feature should be present in the AMPT model with the string melting mechanism owning to the frequent parton-parton scatterings. However, in our model the hadronization mechanism is rather crude (a simple coalescence) thus we should not expect the model to be particularly successful in describing the spectra (in contrast to the studied long-range rapidity correlation which presence or absence is independent on the particular mechanism of hadronization). Nevertheless, it is interesting to investigate whether the AMPT model can approximately reproduce the trends observed in the data. In Fig. \ref{fig:spectra} we present the $p_{T}$ distributions of produced pions, kaons and protons in p+Pb collisions for several centrality classes. The model, despite its simplicity, reproduces the CMS data \cite{Chatrchyan:2013eya} within the accuracy of $20\%$. The calculated spectra shift towards higher $p_t$ with an increasing number of produced particles, $N_{\rm{track}}$, as best visible in the rightmost plot (p+\={p}).

\section{Comments}
\label{sec:comments}

It is worth noticing that the incoherent scattering of partons with basically one essential parameter, $\sigma = 1.5 - 3$ mb, allows to capture the main features of the p+p and p+Pb data for all measured multiplicities and the transverse momenta. This may be contrasted with the CGC framework \cite{Dusling:2013oia} where the saturation scale is fitted separately for each multiplicity and the colliding system.

The presence of the near-side peak in our results originates from the parton scatterings at the early stage of a collision, see Figure \ref{fig:pPb_main_2}. Obviously the lifetime, $\tau$, of the partonic stage increases with increasing number of initial partons, and consequently with $N_{\mathrm{track}}$. We checked that in p+Pb collisions $\tau \sim N_{\mathrm{track}}^{\alpha}$ with $\alpha \sim 1/2$ and for $N_{\mathrm{track}}=50, 100, 200$ the lifetime $\tau \approx 1, 1.4, 1.7$ fm, respectively. In p+p collisions $\tau$ grows slowly from $\tau \approx 0.6$ fm for $N_{\mathrm{track}}=10$ to $\tau \approx 0.8$ fm for $N_{\mathrm{track}}=100$. Our results indicate that for small and rapidly expanding systems there is enough time for multiple parton scatterings which can translate the initial anisotropy of produced matter into the final momentum anisotropy. 

There are several problems in our approach that require further studies. For example only two-to-two elastic parton scatterings are included and higher order processes might become important at high densities. For a complete discussion of various problems in the partonic stage of the AMPT model we refer the reader to Section VII in Ref. \cite{Lin:2004en}. 

A transport model calculations reported in Ref. \cite{Molnar:2004yh} suggest that a parton-parton cross-section of the order of $50$ mb is needed to generate a sizable elliptic flow in A+A collisions. However, in the AMPT model a cross-section of the order of $1.5 - 5$ mb is enough to reproduce the A+A data. It would be interesting to understand the origin of this contradiction.\footnote{We thank D. Molnar and P. Petreczky for comments on this point.}

We would like to emphasize that our goal was not to fit precisely the data. Our objective was to check if a minimal implementation of partonic scatterings, with a reasonable cross-section, can roughly reproduce the experimental data for p+p and p+Pb collisions. As seen in Figure \ref{fig:pPb_main_2}, the agreement with the experimental data is surprisingly good, suggesting that various shortcomings present in our approach are not very important. 

It would be interesting to extend our discussion for peripheral Pb+Pb collisions. We leave this problem for a separate investigation. Also the detailed discussion of the elliptic and triangular \cite{Alver:2010gr} Fourier coefficients will be reported elsewhere.

\section{Conclusions}
\label{sec:conclusions}

In summary, we demonstrated that the incoherent scattering of partons in the
early stage of p+p and p+A collisions is sufficient to understand the
near-side azimuthal correlation of particles separated by a large gap in
pseudorapidity. Using the multi-phase transport model (AMPT with string melting), with a
parton-parton cross-section of $1.5$ mb, we calculated the two-particle
correlation function as a function of $\Delta \eta $ and $\Delta \phi $.
The main trends observed in the data were successfully reproduced. The
near-side peak at $\Delta \phi =0$ is gradually growing with the number of
produced particles owing to the growing density of partons. This in
consequence leads to more frequent parton-parton scatterings. Moreover, the
signal is best visible in the transverse momentum range $1<p_{T}<2$ GeV/$c$,
being in agreement with the CMS data. 

In the default AMPT model, where only partons from minijets interact and soft strings decay independently into particles, the number of interacting partons is not sufficient to produce a visible signal.

Our study indicates that even in a very small system, as the one created in
a p+p collision, there is enough time for partonic scatterings before
the system becomes dilute. These scatterings translate the initial
anisotropy of matter into the final momentum anisotropy, leading to the $\cos
(2\Delta \phi )$ term (and higher harmonics) in the azimuthal correlation
function.

In this paper we focused solely on the main features of the two-particle
correlation function. Calculations of the elliptic and triangular Fourier
coefficients in p+p, p+A and peripheral A+A collisions are left for a
separate investigation.

\section*{Acknowledgments}

We thank Wei Li for clarifications regarding the CMS results.
Discussions with L. McLerran, V. Skokov and R. Venugopalan are appreciated.

G.-L.M. is supported by the Major State Basic Research Development Program
in China under Contract No. 2014CB845404, the NSFC of China under Projects
No. 11175232, No. 11035009, and No. 11375251, the Knowledge Innovation
Program of CAS under Grant No. KJCX2-EW-N01, CCNU-QLPL
Innovation Fund under Grant No. QLPL2011P01, and the ``Shanghai Pujiang
Program'' under Grant No. 13PJ1410600. 

A.B. is supported through the RIKEN-BNL Research Center and the grant No. UMO-2013/09/B/ST2/00497.


\end{document}